\documentclass[prd,nofootinbib,superscriptaddress,twocolumn,preprintnumbers]{revtex4}

\usepackage[mathscr]{eucal}
\usepackage[latin1]{inputenc}
\usepackage{amsmath}
\usepackage{amsfonts}
\usepackage{amssymb}

\usepackage[usenames,dvipsnames]{color}

\usepackage{subfig}
\usepackage{graphicx}
\usepackage{sidecap}

\newcommand{\C}{\mathbb{C}}

\def\id{\mathbb{I}}

\def\H{{\cal H}}

\def\tr{{\rm Tr}}

\def\g{\gamma}

\def\w{{}^o\!\omega}

\newcommand{\f}{\frac}

\usepackage{enumerate}

\usepackage{colordvi}
\newcounter{mnotecount}[section]

\newcommand{\comment}[1]{}
\def\f{\frac}

\def\d{\textrm{d}}

\usepackage{enumerate}

\newcommand{\be}{\nopagebreak[3]\begin{equation}}
\newcommand{\ee}{\end{equation}}
\newcommand{\ba}{\nopagebreak[3]\begin{eqnarray}}
\newcommand{\ea}{\end{eqnarray}}
\newcommand{\nn}{\nonumber \\}

\newcommand{\ket}[1]{\ensuremath{|#1\rangle}}
\newcommand{\bk}[2]{{\langle#1\,|\,#2\rangle}}
\newcommand{\bek}[3]{{\langle#1\,|\,#2\,|\,#3\rangle}}

\def\ini{\textrm{i}}
\def\fin{\textrm{f}}

\begin{document}

\title{Towards Spinfoam Cosmology}

     \author{Eugenio Bianchi}
 \email{bianchi@cpt.univ-mrs.fr}
     \affiliation{Centre de Physique Th\'eorique de Luminy\footnote{Unit\'e mixte de recherche (UMR 6207) du CNRS et des Universit\'es de Provence (Aix-Marseille I), de la M\'editerran\'ee (Aix-Marseille II) et du Sud (Toulon-Var); laboratoire affili\'e \`a la FRUMAM (FR 2291).}
     , Case 907, F-13288 Marseille, EU}
        
      \author{Carlo Rovelli}
 \email{rovelli@cpt.univ-mrs.fr}
     \affiliation{Centre de Physique Th\'eorique de Luminy\footnote{Unit\'e mixte de recherche (UMR 6207) du CNRS et des Universit\'es de Provence (Aix-Marseille I), de la M\'editerran\'ee (Aix-Marseille II) et du Sud (Toulon-Var); laboratoire affili\'e \`a la FRUMAM (FR 2291).}
     , Case 907, F-13288 Marseille, EU}        

     \author{Francesca Vidotto}
 \email{vidotto@cpt.univ-mrs.fr}
     \affiliation{Centre de Physique Th\'eorique de Luminy\footnote{Unit\'e mixte de recherche (UMR 6207) du CNRS et des Universit\'es de Provence (Aix-Marseille I), de la M\'editerran\'ee (Aix-Marseille II) et du Sud (Toulon-Var); laboratoire affili\'e \`a la FRUMAM (FR 2291).}
     , Case 907, F-13288 Marseille, EU}
     \affiliation{Dipartimento di Fisica Nucleare e Teorica,
        Universit\`a degli Studi di Pavia, and\\  Istituto Nazionale
        di Fisica Nucleare, Sezione di Pavia, via A. Bassi 6,
        I-27100 Pavia, EU}
        

\begin{abstract}                 \vskip1em

\noindent We compute the transition amplitude between coherent quantum-states of geometry peaked on homogeneous isotropic metrics.  We use the holomorphic representations of loop quantum gravity and the Kaminski-Kisielowski-Lewandowski generalization of the new vertex, and work at first order in the vertex expansion, second order in the graph (multipole) expansion, and first order in volume${}^{-1}$.  We show that the resulting amplitude is in the kernel of a differential operator whose classical limit is the canonical hamiltonian of a Friedmann-Robertson-Walker cosmology. This result is an indication that the dynamics of loop quantum gravity defined by the new vertex yields the Friedmann equation in the appropriate limit. 

\end{abstract}

\maketitle


\section{Introduction}

The dynamics of loop quantum gravity (LQG) can be given in covariant form by using the {\em spinfoam} formalism.  In this paper we apply this formalism to cosmology. In other words, we introduce a spinfoam formulation of quantum cosmology, or a ``spinfoam cosmology".  We obtain two results.  The first is that physical transition amplitudes can be computed, in an appropriate expansion.  We compute explicitly the transition amplitude between homogeneous isotropic coherent states, at first order.  The second and main result is that this amplitude is in the kernel of an operator $\hat C$, and the classical limit of $\hat C$ turns out to be precisely the Hamiltonian constraint of the Friedmann dynamics of homogeneous isotropic cosmology.  In other words, we show that LQG yields the Friedmann equation in a suitable limit.  

LQG has seen momentous developments in the last few years.  We make use of several of these developments here, combining them together. The first ingredient we utilize is the ``new" spinfoam vertex \cite{Engle:2007uq,Livine:2007vk,Engle:2007qf,Freidel:2007py,Engle:2007wy}. The second is the Kaminski-Kisielowski-Lewandowski extension of this to vertices of arbitrary-valence \cite{Kaminski:2009fm}. The third ingredient is the coherent state technology 
\cite{Hall:2002,
Ashtekar:1994nx,
Thiemann:2000bw,Thiemann:2000ca,Thiemann:2000bx, Thiemann:2000by,Sahlmann:2001nv,Thiemann:2002vj,
Bahr:2007xa,Bahr:2007xn,Flori:2008nw,Flori:2009rw,
Freidel:2010aq,Freidel:2010}, and in particular the holomorphic coherent states discussed in detailed in \cite{Bianchi:2009ky}. These states define a holomorphic representation of LQG \cite{Ashtekar:1994nx,Bianchi:2010}, and we work here in this representation. 

Our strategy is the following. We consider the standard Hilbert space of canonical LQG and we assume the dynamics to be given by the new vertex.  We consider holomorphic coherent states in this Hilbert space and we work in the holomorphic representation they define. 

We truncate LQG down to a graph with a finite number of links. In particular, the calculation is based on the ``dipole"  graph formed by two nodes connected by four links
\cite{Rovelli:2008dx}.  This choice determines a Hilbert space, which describes a finite number of the degrees of freedom of the gravitational field. These degrees of freedom can be identified as the lowest modes in a multipole expansion of the metric in hyper-spherical harmonics on $S_3$ \cite{Battisti:2009kp}.  That is, they describe a closed cosmology, with anisotropies and a few low-mode inhomogeneities.   

In particular, we consider coherent states that are peaked on homogeneous isotropic geometries.  We emphasize the fact that these states are just peaked on homogeneous and isotropic geometries, but they also include fluctuations of the inhomogeneous and anisotropic degrees of freedom. So, the dynamics of the quantum theory we consider {\em does} include inhomogeneous and anisotropic degrees of freedom.  Homogeneous-isotropic coherent states are labelled by two parameters which capture the scale factor $a$ of standard cosmology and its time derivative $\dot a$; or, equivalently, the $p$ and $c$ canonical variables used in Loop Quantum Cosmology (LQC). In the holomorphic representation, these two quantities appear in the complex combination $z=\alpha c+i\beta p$, and therefore the states we consider are labelled by the complex number $z$. The transition amplitude between two such states is then an analytic function $W(z,z')$ of two complex variables. 

We write this transition amplitude at first order in a vertex expansion. We view this as the analog of a first order calculation in, say, QED perturbation theory. We compute explicitly $W(z,z')$ in the limit in which the geometry is large compared to the Planck scale.  

In other words, we compute the transition amplitude between macroscopical homogeneous isotropic cosmological spaces in LQG, taking three approximations from the complete theory: (i) the truncation of the degrees of freedom to those defined on a finite graph; (ii) the restriction to first order in the vertex expansion; (iii) the large volume limit.  The validity of these approximation can only be justified a posteriori, from the correctness of the result. 

Our next step is to notice that the transition amplitude computed solves the equations \mbox{$H\, W(z,z')=0$} for a certain operator $H=H(z, {\scriptstyle\hbar} \frac{\d}{\d z})$. This fact implies that the amplitude defines a quantum dynamics where the operator constraint $H=0$ holds.  The corresponding classical dynamics will be governed by (the $\hbar\to 0$ limit of) the classical constraint $H(z, \bar z)=0$. When written in terms of $p$ and $c$, this turns out to be precisely the Hamiltonian constraint that governs (the gravitational part of) the dynamics of a classical Friedmann cosmology, in the limit of large volume.  Therefore LQG yields the Friedmann dynamics in this limit. 

Several words of caution are necessary.  First, we work in the Euclidean theory. Second, the cosmological dynamics that we obtain is the one in the large volume limit and since we do not have any matter present, this has only the solution $a=constant$, which is flat space.  With these caveats, our result is that there is an approximation in LQG that leads to classical cosmology. 

This result can be compared with those of LQC \cite{Bojowald:2006da,Ashtekar:2008zu}. In LQC, one first reduces the classical theory to a cosmological system with a finite number of degrees of freedom, and then applies a ``loop quantization" to this symmetry reduced model. Thus, one has a complete quantum theory of a truncation of the classical theory.  Here, instead, we start from the full quantum theory and take an approximation. Therefore in LQC one studies exact solution in a truncated system, while here we study approximated solutions in the (hopefully) exact quantum theory.   The possibility of introducing a spinfoam-like expansion starting from LQC has been considered in the papers \cite{Ashtekar:2009dn,Ashtekar:2010ve,Rovelli:2009tp}. These papers and the present work can be seen as two converging attempts to construct a spinfoam version of quantum cosmology. 

Finally, in our opinion a main reason of interest of the result we present here is that it represents an example of a complete calculation of physical transition amplitude in background independent quantum gravity. It complements the calculation of the two-point function \cite{Rovelli:2005yj,Bianchi:2006uf,Alesci:2007tx,Alesci:2007tg,Alesci:2008ff}, that has been recently completed \cite{Bianchi:2009ri}.

\vskip.5cm

In Section \ref{theory}, we briefly recall the definition of the full quantum theory. In 
Section \ref{cosmo}, we discuss the approximation that selects a cosmological sector, we compute the resulting transition amplitude. In Section \ref{limit}, we study the classical limit and recover the Friedmann dynamics. 
 
\section{The theory}\label{theory} 

\subsection{Kinematics}

The theory is defined on the Hilbert space
\be
     {\cal H}= \bigoplus_\Gamma\ {\cal H}_\Gamma
\ee
The sum runs over the abstract graphs $\Gamma$. A graph $\Gamma$ is here a set of $L$ links $\ell$ and $N$ nodes $n$, together with two relations $s$ (source) and $t$ (target) assigning a source node $s(\ell)$ and a target node $t(\ell)$ to every link $\ell$. 
The Hilbert space $ {\cal H}_\Gamma$ is defined to be 
\be
 {\cal H}_\Gamma= L_2[SU(2)^L/SU(2)^N]
\ee
where the action of $SU(2)^N$ on the states \mbox{$\psi(U_l)\in L_2[SU(2)^L]$} \, is 
\be
\psi(U_l) \to \psi(V_{s(\ell)}U_l V_{t(\ell)}^{-1})~,\hspace{2em} V_n\in SU(2)^N\,.
\label{gauge}
\ee

Two sets of operators are defined on each space $\H_{\Gamma}$. For each link we have the ``holonomy" multiplicative operator $\hat U_\ell\,\psi(U_\ell)=U_\ell\,\psi(U_\ell)$ and the ``triad" operator $\hat E^i_\ell\,\psi(U_\ell)=(8\pi G\hbar\gamma) \,L^i_\ell\,\psi(U_\ell)$ where $G$ is the Newton constant and $L^i_\ell$ is the left-invariant vector field acting on the variable $U_\ell$.

Finally, the state space of the theory is obtained by factoring $\H$ by an equivalence relation,  defined as follows.  If $\Gamma$ is a subgraph of $\Gamma'$ then ${\cal H}_\Gamma$ can be naturally identified with a subspace of  ${\cal H}_{\Gamma'}$. Two states are equivalent if they can be related by this identification, or if they are mapped into each other by the discrete group of the automorphisms of $\Gamma$.

These states are to be thought as ``boundary states" in the quantum theory. That is, $\H$ must be identified with the space $\H^*_{out}\otimes \H_{in}$ of the initial and final states of non-relativistic quantum mechanics. 

\subsubsection*{Coherent states}

An overcomplete basis of coherent states in the Hilbert space ${\cal H}_\Gamma$ is provided by the holomorphic states 
\be
\Psi_{H_\ell}(U_\ell)=\int_{SU(2)^N}  \d g_n   \prod_{\ell} K_t(g_{s(\ell)}^{-1}U_\ell  \,  \, g_{t(\ell)}\, H_{\ell}^{-1}\, ).
\label{states}
\ee
Here $H_\ell\in SL(2,\C)$, and $K_t$ is (the analytic continuation to $SL(2,\C)$ of) the 
 \emph{heat kernel} function on $SU(2)$. This is a function concentrated on the origin of the group, with a spread of order~$t$. Its explicit form is\footnote{We choose a parameter $t$ with the dimension of an inverse action, and put $\hbar$ explicitly in the definition of the coherent states, in order to emphasize the fact that the small $t$ limit is the classical limit, and to keep track of the corresponding dependence on $\hbar$. The factor 2 is for later convenience.} 
\be
K_t(U) = \sum_j (2j+1)  e^{-2 t\hbar\, j(j+1)} \   \tr[D^j(U)]
\ee
where $D^j(U)$ is the Wigner matrix of the spin-$j$ representation of $SU(2)$. 

As shown in  \cite{Bianchi:2009ky}, these states: (i) are the basis of the holomorphic representation \cite{Ashtekar:1994nx,Bianchi:2010}, (ii) are a special case of Thiemann's complexifier's coherent states 
\cite{Thiemann:2000bw,Thiemann:2000ca,Thiemann:2000bx, Thiemann:2000by,Sahlmann:2001nv,Thiemann:2002vj,
Bahr:2007xa,Bahr:2007xn,Flori:2008nw,Flori:2009rw},
(iii) induce Speziale-Livine coherent tetrahedra \cite{Livine:2007vk,Conrady:2009px,Freidel:2009nu} on the nodes, and (iv) are equal to the the Freidel-Speziale coherent states \cite{Freidel:2010aq,Freidel:2010} for large spins.

The states \eqref{states} are gauge-invariant semiclassical wave packets.  The integral in 
\eqref{states} projects (``group averages") on the gauge invariant states. If $H_\ell$ is in the $SU(2)$ subgroup of $SL(2,\C)$, the heat kernel peaks each $U_\ell$ on $H_\ell$. The extension of $H_\ell$ to $SL(2,\C)$ has the same effect as taking a gaussian function $\psi(x)=e^{(x-z_o)^2/2}\sim e^{(x-x_o)^2/2} e^{ip_ox}$ for a complex $z_o=x_o+ip_o$; that is, it adds a phase which peaks the states on a value of the variable conjugate to $U_\ell$. Thus, the states \eqref{states} are peaked on the variables $U_\ell$ as well as on their conjugate momenta. 

The $SL(2,\C)$ labels $H_\ell$ can be given two related interpretations. First, we can decompose each $SL(2,\C)$ label in the form
\be
         H_\ell=e^{i4tE_\ell/8\pi G\gamma}\ U_\ell
         \label{heu}
\ee 
where $U_\ell\in SU(2)$ and $2tE_\ell/(8\pi G\hbar\gamma)\in su(2)$.  Then it is not hard to show that $U_\ell$  and $E_\ell$ determine the expectation value of the operators  $\hat U_\ell$ and $\hat E_\ell$ on the state $\psi_{H_\ell}$
\be
         \frac{  \bek{\psi_{H_\ell}}{\hat U_\ell}{\psi_{H_\ell}}  }{  \bk{\psi_{H_\ell}}{\psi_{H_\ell}} }=U_\ell~, \hspace{1em}
         \frac{  \bek{\psi_{H_\ell}}{\hat E_\ell}{\psi_{H_\ell}}  }{  \bk{\psi_{H_\ell}}{\psi_{H_\ell}} }= E_\ell~,
\ee 
and that the corresponding spread is small\footnote{If we fix a length scale $L\gg \sqrt{\hbar G}$ and choose $2\hbar t=\hbar G/L^2$ we have $
        \Delta U_\ell \sim ~ \sqrt{\hbar G}/L$ and $
        \Delta E_\ell \sim \sqrt{\hbar G}\,L$.
}
\be
        \Delta U_\ell \sim ~ \sqrt{t\hbar} , \hspace{1.5em}
        \Delta E_\ell \sim G\sqrt{\hbar/t}. 
\ee 

Alternatively, we can decompose each $SL(2,\C)$ label in the form
\be
H_\ell = n_{s,\ell} ~ e^{-i(\xi_\ell+i\eta_\ell)\f{\sigma_3}{2}} ~ n^{-1}_{t,\ell}~; 
\label{fs}
\ee
$\vec \sigma=\{\sigma_i\}, i=1,2,3$ are the Pauli matrices. Freidel and Speziale discuss a compelling geometrical interpretation for the $(\vec n_s, {\vec n}_t, \xi, \eta)$ labels of each link  \cite{Freidel:2010aq}.  For appropriate four-valent states representing a Regge 3-geometry with intrinsic and extrinsic curvature, the vectors ${\vec n}_s, {\vec n}_t$ are the $3d$ normals to the triangles the tetrahedra bounded by the triangle; $\xi$ is the extrinsic curvature at the triangle and $\eta$ is the area of the triangle divided by $8\pi G\hbar$.  For general states, the interpretation extends to a simple generalization of Regge geometries, that Freidel and Speziale have baptized ``twisted geometries". 

Thus, the holomorphic coherent states provide a convenient basis of wave packets with  good geometrical interpretation. 
       
\subsection{Dynamics} 

The spinfoam formalism associates an amplitude
\be
\bk W\psi=\sum_\sigma \ \prod_f d_f(\sigma)   \prod_v W_v(\sigma)
\label{sf}
\ee
to each boundary state $\psi\in\H$. The sum is over the spinfoams $\sigma$ bounded by $\psi$.  See \cite{Engle:2007wy} for a description of this formalism and for the notation. 
The vertex amplitude is $W_v(\sigma) =  \langle W_v| \psi_v(\sigma)\rangle$, 
where $\psi_v(\sigma)$ is the spin network obtained by cutting $\sigma$ with a small 3-sphere surrounding $v$ and the the vertex amplitude that defines the dynamics of LQG is  \cite{Engle:2007wy,Engle:2007wy,Livine:2007vk,Freidel:2007py,Kaminski:2009fm} 
\be
  \langle W_v| \psi\rangle =   (f\psi)(\id)~.
\label{vertex}
\ee
Here $f:\H_\Gamma\to 
L_2[SO(4)^L/SO(4)^N]$ is defined as follows.  Let
\be
j^\pm=\frac{1\pm\gamma}{2} j
\ee
and let $Y$ be the map
\ba
Y:  ~~~  \H^{(j)} &~~\longrightarrow~~&  \H^{(j^+,j^-)}  \nn
       \ket{j,m} && \ket{j^+,m^+;j^-,m^-}
\ea
whose matrix elements are given by the Clebsh-Gordan coefficients. 
\be
Y_m^{m^+ m^-}
= \bk{j^+,m^+;j^-,m^-}{j,m} ~.
\ee 
Consider the Peter-Weyl decomposition of $L_2[SU(2)^L]$ and, respectively  $L_2[(SO(4)^L]$. Then $f$ is defined by mapping  with $Y$ each $\H_j$ terms of the first into the corresponding $\H_{j^+,j^-}$ of the second.

Explicitly, the generalized state $W_v$ in \eqref{vertex} is given by 
\be
W_v(U_{\ell})  =
\int_{SO(4)^N}  
\d G_n \, ~
\prod_{\ell}    P_o(U_{\ell} \, , \,G_{s(\ell)} G_{t(\ell)}^{-1} )
\label{wu}
\ee
where
\be
P_o(U,G)\! =\! 
\sum_{j} {\scriptstyle (2j+1)}\ \tr\!
\left[
D^{\scriptscriptstyle(j)}(U)\, Y^\dagger \, D^{\scriptscriptstyle(j^{\!+}\!\!,j^{\!-}\!)}\!(G) \,Y
\right]\!.
\ee
Here $D^{(j)}$ is the Wigner matrix of the spin-$j$ representation of $SU(2)$, while $D^{(j^{\!+}\!,j^{\!-}\!)}$ is the Wigner matrix of the spin-$(j^{\!+}\!,j^{\!-}\!)$ representation of $SO(4)$. The first has has dimension $2j+1$ while the second has dimension $(2j^++1)(2j^-+1)$. These matrices with different dimensions are glued by the map $Y$.

The vertex amplitude takes a simple form on the holomorphic basis defined by the coherent states. By combining the definition \eqref{wu} of the vertex and the definition \eqref{states} of the coherent states, one obtains the holomorphic form of the vertex amplitude \cite{Bianchi:2010}
 \ba 
W_v(H_{\ell})&\!\equiv&\! \bk{W_v}{\psi_{H_\ell}}=
\int_{SU(2)^L} \!\! \d U_\ell \ W(U_\ell)\  \psi_{H_\ell}(U_\ell) \nn
&=&\int_{SO(4)^N}  \d G_n \, ~
\prod_{\ell}     P_t(H_{\ell} \, , \,G_{s(\ell)} G_{t(\ell)}^{-1} )
\label{daniele}
\ea
where 
\be
P_t(H,G)\! =\! \! 
\sum_{j} {\scriptstyle (2j+1)}e^{\scriptscriptstyle-2t\hbar j(j+1)} \tr\!
\left[\! 
D^{\scriptscriptstyle(j)}\!(H)Y^\dagger D^{\scriptscriptstyle(j^{\!+}\!\!,j^{\!-}\!)}\!(G)Y
\! \right]\!.
\label{daniele2}
\ee
Here $D^{(j)}$ is the analytic continuation of the Wigner matrix from $SU(2)$ to $SL(2,\C)$.  Below, we use this last expression to compute the quantum evolution in cosmology. 

\section{The cosmological approximation}\label{cosmo}

\subsection{Graph expansion}

There is no physics without approximations. The first approximation we take is to truncate $\H$ to a single fixed graph $\Gamma$.  Notice that the states with support on smaller graphs (subgraphs of $\Gamma$) are all contained in $\H_\Gamma$; therefore the truncation amounts to disregard all states that have support on graphs ``larger" that $\Gamma$. 

We choose $\Gamma$ to be the graph formed by two disconnected components $\Gamma_{\ini}$ and $\Gamma_\fin$.  We think at these as carrying an initial and a final state.  In particular we choose   $\Gamma_{\ini}=\Gamma_\fin=\Delta_2^*$, where the ``dipole" graph $\Delta_2^*$ is defined by the set of two nodes $\{n_1,n_2\}$, by the set of four links $\{\ell_1,\ell_2,\ell_3,\ell_4\}$, and by the source and target relations $s(\ell)=n_1$ and  $t(\ell)=n_2, \forall \ell$. That is\\[-1.5mm]
\begin{center}
\begin{picture}(20,40)
\put(-36,28) {$\Delta_2^*\  = $}
\put(04,30) {\circle*{3}} 
\put(36,30) {\circle*{3}}  
\qbezier(4,30)(20,53)(36,30)
\qbezier(4,30)(20,21)(36,30)
\qbezier(4,30)(20,39)(36,30)
\qbezier(4,30)(20,7)(36,30)
\put(50,27) {\circle*{1}} 
\end{picture}
\end{center} 
\vspace{-1.5em}%
The operators defined on the Hilbert space $\H_{\Delta_2^*}$ are $(U_\ell,E_\ell)$. These can be interpreted as follows. Consider a space $M$ with the topology of a three sphere, carrying a triad field $E$ and a connection $A$. Choose an immersion of $\Delta_2^*$ into $M$, and a cellular decomposition $\Delta_2$ of $M$, dual to $\Delta_2^*$.  $\Delta_2$ is the triangulation of the 3-sphere formed by 2 tetrahedra with all their faces identified.\footnote{The metric structure defined by $E$ determines a preferred immersion, up to degeneracies. Pick two points $n_1$ and $n_2$ at maximal distance from each other (the ``north" and ``south" pole in $M$), and let the equator be the set of points equidistant from the poles. Chose four points on the equator at maximal distance from one another, and connect them to the poles with geodesic links $\ell$ (this gives the four meridians). The ``equator" gets partitioned into four (Vorono\"i) triangles \cite{Voronoi}, each cut by one of the links, defined by the minimal distance from the cuts.\label{foot1}}  We can then identify $U_\ell$ with the holonomy of $A$ on the link $\ell$ and $E_\ell$ with the flux of the triad through the triangle cut by the link $\ell$, parallel transported to $n_1$.

    \subsubsection*{Homogeneous isotropic coherent states}
    
Consider the coherent states on     $\H_{\Delta_2^*}$. These are labelled by four $SL(2,\C)$ elements $H_\ell=e^{iE_\ell} U_\ell$. We are interested here in 
homogeneous isotropic coherent states.  To find them, let us compute  $(U_\ell,E_\ell)$ for the case of a homogeneous isotropic space. Let $A$ and $E$ to define such a space. Then we can write $A= c\ \w$ and $E= p\ \w$ where $\w=g^{-1}dg$
is the fiducial connection defined by the $SU(2)$ Maurer-Cartan connection, upon identification of $M$ with the $SU(2)$ group manifold \cite{Ashtekar:2003hd}.  Choose $n_1$ to be the identity $\id$ and $n_2$ to be $-\id$, and choose the links to be given by exponentiating a quadruplet $\vec n_\ell$ of vectors in $su(2)$ normal to the faces of a regular tetrahedron centered on the origin.   There is an $SO(3)$ freedom in choosing the normals at each node. We choose $n_{1\,\ell}=n_{2\,\ell}:=n_\ell$.
 This gives a realization of the immersion defined in the footnote \ref{foot1}. Then we can compute $U_\ell$ and $E_\ell$ \cite{Battisti:2009kp}
\be
  U_\ell=n_\ell ~ e^{-i \alpha  c\f{\sigma_3}{2}} ~ n^{-1}_\ell~,\hspace{1em}
  E_\ell=-i n_\ell ~ \frac{2 \pi G\g}t \beta p\f{\sigma_3}{2} ~ n^{-1}_\ell
\ee
where $n_\ell$ are $SU(2)$ group elements such that $n_\ell\sigma^3 n^{-1}_\ell=\vec n_\ell\cdot \vec \sigma$, and $\alpha$ and $\beta$ are constants that we do not determine here. This implies that in \eqref{fs} we have $n_{s,\ell}=n_{t,\ell}=n_{\ell}$ and
\be
\xi_\ell=\xi=\alpha c, \hspace{1em}
\eta_\ell=\eta=\beta p~,
\label{ab}
\ee
that is 
\be
H_\ell(\xi,\eta) = n_{\ell} ~ e^{-i(\xi+i\eta)\f{\sigma_3}{2}} ~ n^{-1}_{\ell}~.
\label{fs2}
\ee
The independence of $\xi_\ell$ and $\eta_\ell$ from $\ell$ can be seen as the effect of isotropy and the equality of of $n_{s,\ell}$ and $n_{t,\ell}$ can be seen as the effect of homogeneity.  The two numbers $c=\xi/\alpha$ and $p=\eta/\beta$ label the homogeneous isotropic coherent states. 

Remarkably, the same states can be obtained by using the Friedel-Speziale geometrical interpretation \cite{Freidel:2010}. Consider a Regge geometry formed by two equal regular tetrahedra with their faces identified and where the extrinsic curvature is the same at each triangle.  This fixes the labels $(n_s, n_t, \xi, \eta)$ at each triangle $\ell$, and determines via \eqref{fs} an $SL(2,\C)$ group element  which is precisely \eqref{fs2}. 

Finally, the quantity that we want to calculate is
\be
W(\xi_{\ini},\eta_{\ini};\xi_{\fin},\eta_{\fin}) = W(H_l(\xi_{\ini},\eta_{\ini}),H_l(\xi_{\fin},\eta_{\fin}))
\label{questo}
\ee
Notice that this is an holomorphic function of $z_{\ini}$ and $z_{\fin}$ where
\be 
z \equiv \xi+i\, \eta~.
\label{z}
\ee
Thus we can write it as 
\be
W(z_{\ini},z_{\fin}) = W(\xi_{\ini},\eta_{\ini};\xi_{\fin},\eta_{\fin})~.
\ee
This is the transition amplitude 
 between a homogeneous isotropic universe with scale factor (square) $p_{\ini}$ and extrinsic curvature $c_{\ini}$, to a a homogeneous isotropic universe with scale factor (square) $p_{\fin}$ and extrinsic curvature $c_{\fin}$. 

By writing the in and out states 
\be
\label{def_states}
\psi_{z}(U_\ell) :=\psi_{H_\ell(z(c,p))} (U_\ell) := \bk{U_\ell}{z},
\ee
we can interpret $W(z_{\ini},z_{\fin})$ as the physical scalar product between (the projection on the physical state space of) the state $\ket z_{\ini}$ and  (the projection on the physical state space of) the state $\ket z_{\fin}$. That is\footnote
{In standard quantum mechanics, the transition amplitude $Z(z_\fin,  z_\ini) = \bek{z_\fin}{\exp{iHt}}{z_\ini}$ is anti-linear in the first variable, therefore one may expect 
\eqref{Wzz} to be anti-holomorphic in $z_\fin$. However, since here we treat the past and future surfaces as two components of the boundary of the spacetime region between the two, the initial surface is oriented the towards the future and the final surface towards the past. If we change the orientation of the final surface, $z_\fin$ goes to  $-\bar z_\fin$. 
}
\be
W(z_{\ini}, z_{\fin}) = \bk{\bar z_\fin}{z_\ini}_{\rm physical}
\label{Wzz}
\ee
We now compute this quantity, to first order in the vertex expansion.

\subsection{Vertex expansion}

At first order in the vertex expansion, the amplitude \eqref{Wzz} is given by the spinfoam formed by a single vertex bounded by four edges and eight faces.
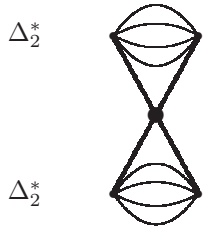
\begin{center}
\begin{picture}(20,40)
\put(-36,28) {$\Delta_2^*\  $}
\put(04,30) {\circle*{3}} 
\put(36,30) {\circle*{3}}  
\qbezier(4,30)(20,53)(36,30)
\qbezier(4,30)(20,21)(36,30)
\qbezier(4,30)(20,39)(36,30)
\qbezier(4,30)(20,7)(36,30)
\put(-36,-32) {$\Delta_2^*\  $}
\put(04,-30) {\circle*{3}} 
\put(36,-30) {\circle*{3}}  
\qbezier(4,-30)(20,-53)(36,-30)
\qbezier(4,-30)(20,-21)(36,-30)
\qbezier(4,-30)(20,-39)(36,-30)
\qbezier(4,-30)(20,-07)(36,-30)
\linethickness{0.4mm}
\put(20,0) {\circle*{6}} 
\qbezier(4,29)(20,0)(36,-29)
\qbezier(4,-29)(20,0)(36,29)
\end{picture} 
\end{center} 
\vskip13mm
 in this approximation \eqref{questo} is given by the amplitude of the single vertex $v$
\be
W(z_{\ini},z_{\fin}) = W_v(H_\ell(z_\ini),H_\ell(z_\fin))
\ee
Using (\ref{daniele},\ref{daniele2}) this becomes
\ba
 W(z_{\ini},z_{\fin})  &=& 
\int_{SO(4)^4}  
\d G_1^{\ini} \,\d G_2^{\ini} \,\d G_1^{\fin} \,\d G_2^{\fin} \, ~\times\nn&&\!\!\!\!\!\!\!\!\!\!\!\!\!\!\!\!\!
\prod_{\ell^{\ini}}       P_t  (H_{ \, \ell}(z_\ini) \, , \,G_1^{\ini} G_2^{\ini \, -1} )
\prod_{\ell^{\fin}}     P_t  (H_{ \, \ell}(z_\fin) \, , \, G_1^{\fin} G_2^{\fin \, -1} )
\nn &=& 
\int_{SO(4)^2}  
\d G^{\ini} \,\d G^{\fin} \, ~\times\nn&&\!\!\!\!\!\!\!\!\!\!\!\!\!\!\!\!\!
\prod_{\ell^{\ini}}       P_t  (H_{ \, \ell}(z_\ini) \, , \,G^{\ini}  )
\prod_{\ell^{\fin}}     P_t  (H_{ \, \ell}(z_\fin) \, , \, G^{\fin} )
\nonumber
\ea
Notice that this expression factorizes
\be
W(z_{\ini},z_{\fin}) = W(z_{\ini})\ W(z_{\fin})
\label{Wzz2}
\ee
where
\be
 W(z)  =
\int_{SO(4)}  
\d G\ 
\prod_{\ell}       P_t  (H_{ \, \ell}(z) , \,G )
\label{Wz}
\ee 
This factorization happens only at first order.  At this order, therefore, the physical projector projects on a single state, and \eqref{Wz} can be viewed as a Hartle-Hawking 
\emph{no-boundary} state \cite{Hartle:1983ai,Hartle:2008ng}.

\subsection{Large volume expansion}

Let us now compute \eqref{Wz} in the limit in which the universe is large. This limit is given by taking large $p$ in \eqref{heu}. Consider the factor $D^{\scriptscriptstyle(j)}(H)$ in \eqref{daniele2}. Using \eqref{heu}, this reads 
\be
D^{\scriptscriptstyle(j)}(H_\ell)=
D^{\scriptscriptstyle(j)}(n_\ell)\ 
D^{\scriptscriptstyle(j)}(e^{-iz\f{\sigma_3}2})\ 
D^{\scriptscriptstyle(j)}(n^{-1}_\ell)~.
\ee
For $p\gg 8\pi G\hbar\gamma $, namely $\eta\gg 1$ we have
\be
D^{\scriptscriptstyle(j)}(e^{-i(\xi+i\eta)\f{\sigma_3}2})= e^{-i(\xi+i\eta)j}\ P~.
\ee
where $P$ is the projector on the eigenstate of $L_3$ with maximum eigenvalue $m=j$. It is easy to see that this terms dominates in the limit \cite{Bianchi:2010}. Inserting this result in the previous equation gives
\be
D^{\scriptscriptstyle(j)}(H_\ell)= e^{-i zj}\ 
D^{\scriptscriptstyle(j)}(n_\ell)\ 
P\ D^{\scriptscriptstyle(j)}(n^{-1}_\ell):= e^{- i zj}\ P_\ell~.
\ee
Using this in \eqref{daniele2} gives
\be
P_t(H_\ell,G)\! =\! \! 
\sum_{j} {\scriptstyle (2j+1)} e^{\scriptscriptstyle-2t\hbar j(j+1)- i zj}  \tr\!
\left[
P_\ell Y^\dagger  D^{\scriptscriptstyle(j^{\!+}\!\!,j^{\!-}\!)}\!(G) Y\!
\right]\!.\nonumber
\ee
We show later that the trace gives a contribution polinomial in $j$. Therefore we can compute the sum by approximating it with a Gaussian integral. This is peaked on the value $j\sim j_o=-i z/4t\hbar$. Notice that the real part of $j_o$ is given by the imaginary part of $z$, namely $p$. For large $p$, we have $j_o\sim \beta p/4t\hbar$. The gaussian integral gives
\be
P_t(H_\ell,G)= \sqrt{\f\pi t}\ e^{-\f{z^2}{8t\hbar}}\ 2j_o  \tr\!
\left[
P_\ell\, Y^\dagger \, D^{\scriptscriptstyle(j_o^{\!+}\!\!,j_o^{\!-}\!)}\!(G) \,Y
\right]\!.
\ee
Using this in \eqref{Wz} yields
\be
W(z)= \left(\sqrt{{\f\pi t}}\ e^{-\f{z^2}{8t\hbar}}\ 2j_o  \right)^{\!\!4}\, N_j
\ee
where 
\be
N_{j_o} = \int_{SO(4)}  
\d G\ \ 
\prod_{\ell} \    \tr\!
\left[P_\ell\, Y^\dagger \, D^{\scriptscriptstyle(j_o^{\!+}\!\!,j_o^{\!-}\!)}\!(G) \,Y
\right]\!.
\ee
This is norm squared of the Livine-Speziale coherent regular tetrahedron of size $j_o$. It is given \cite{Livine:2007vk} by $N_{j_o}=N_{o}j_o^{-3}$. Using this and $j_o=-\f{i}{4t\hbar} z$, we conclude
\be
W(z)= N z e^{-\f{z^2}{2t\hbar}}
\ee
where $N=-2i(\pi/t\hbar)^{2}N_o$. Finally, inserting into \eqref{Wzz2} we have
\be
W(z_\ini,z_\fin)= N^2\  z_\ini\, z_\fin\ e^{-\f{1}{2t\hbar}(z_\ini^2+z^2_\fin)}~.
\label{eccola}
\ee
This is the transition amplitude between two cosmological homogeneous isotropic coherent states. 

\section{Classical limit}\label{limit}

We now observe that the transition amplitude \eqref{eccola} satisfies the equation  
\be
 \f3{8\pi G(4\alpha\beta\gamma)^2}
\left(z^2-t^2\hbar^2 \f{\d^2}{\d z^2}-3t\hbar\right)^{\!\!2}\, W(z,z')=0
\ee
(the square and the overall factor are for later convenience). Therefore this amplitude describes a quantum system where the operator equation 
\be
    \hat H :=  \f3{8\pi G(4\alpha\beta\gamma)^2}
    \left(z^2-t^2\hbar^2 \f{\d^2}{\d z^2}-3t\hbar\right)^{\!\!2} =0
\label{H}
\ee
holds. Let us now look for a corresponding classical system. Assume that $z$ is the coordinate of a classical phase space with symplectic structure
\be
    \omega= \f{i}{t}\ dz\wedge d\bar z ~;
    \label{ss}
\ee
that is
\be
    \{z,\bar z\} = i t~.
\ee
The corresponding operators in the quantum theory, that satisfy $[z,\bar z]=i\hbar \{z,\bar z\}$ are therefore
\be
\hat z = z~,   ~~~~~ \hat{\bar z} = t\hbar \f\d{\d z}~.
\ee
We can thus rewrite \eqref{H} as
\be
 \hat H =  \f3{8\pi G(4\alpha\beta\gamma)^2}
 \left(\hat z^2-\hat{\bar z}^2 -2 \right)^{\!2} = 0
\ee
Let us now take the classical limit of this equation.  Replacing operators with classical variables, we have 
\ba
 H &=&  \f3{8\pi G(4\alpha\beta\gamma)^2}
 \left(z^2-{\bar z}^2 -2 \right)^{\!2} \nn&=& 
   \f3{8\pi G(4\alpha\beta\gamma)^2}
  \left(4i\xi\eta -2 \right)^{\!2} = 0
\ea
Using \eqref{ab}, this is
\be
 H =  \f3{8\pi G(4\alpha\beta\gamma)^2}
 \left(4i\,\alpha\beta\,cp  -2\hbar \right)^{\!\!2} = 0~.
\ee
In the large $p$  limit we have
\be
 H =- \f{3}{8\pi G\gamma^2}\ c^2p^2= 0~.
\ee
Dividing by the volume of space $V\!ol\sim p^{3/2}>0$, we have
\be
 H =- \f3{8\pi G\gamma^2}\  \sqrt{p}c^2= 0~.
 \label{fine}
\ee
In the large volume limit and in the absence of matter the dynamics of the universe approaches the flat ($k=0$) case and the Friedmann hamiltonian constraint becomes
\be
\label{hclk} H_{cl} = -
\f{3}{8\pi G  } \  \dot a^2 a = 0~;
\ee
in this regime $c=\g\dot a$ and $p=a^2$ and this equation is precisely \eqref{fine}.

\section{Conclusion}

We have introduced a spinfoam formulation of quantum cosmology.

We have obtained two results. The first is that it is possible to compute quantum transition amplitudes explicitly in suitable approximations. In detail, we have studied three approximations: (i) cutting the theory to a finite dimensional graph (the dipole), (ii) cutting the spinfoam expansion to just one term with a single vertex and (iii) the large volume limit.  The main hypothesis on which this work is based is that the regime of validity in which these approximations are viable includes the semiclassical limit of the dynamics of large wavelengths. ``Large" means here of the order to the size of the universe itself.  This regime includes of course the standard Friedmann cosmology.  

The second result is that the transition amplitude computed appears to give the correct Friedmann dynamics in the classical limit.   This results must be taken with caution, for a number of reasons.  First, we have used the Euclidean theory, instead than the physical Lorentzian theory. Second, the dynamics we have obtained is in fact rather trivial.  The solution of the constraint equation \eqref{fine} is either $p=0$ or $c=0$; that is, either the universe has no volume, or it is flat. This is physically correct, since in absence of a matter and in the limit of infinitely large radius one obtain precisely a flat spacetime.  But is is only a weak indication that the full Friedmann dynamics is recovered.  Wether the result still holds with matter, or a cosmological constant, must still be check.  Also, in the derivation of the classical limit, the symplectic structure \eqref{ss} has been taken as an input. It can be shown that this choice reproduces
 the symplectic structures of the LQC variables\footnote{
This fixes the product $\alpha\beta=3t/16\pi G\g$. Then $\alpha$ can be determined by noticing that for $c=1$ the connection $A$ is the Cartan connection and its holonomy from the identity to $g$ is $g$ itself. Taking $n_s=\id$ and $n_t=-\id$ gives easily $\alpha=2\pi$. 
 }
  $(c,p)$ or the one of the LQG variables $(E_\ell,U_\ell)$.  Finally, the system that the approximation defines admits obvious improvements. In particular, transitions must be computed on a larger graph, and at the next order in the vertex amplitude, in order to investigate the validity of the approximation. 

We have noticed that at first order the transition amplitude factorizes (see eq.\eqref{Wzz}). To this order, the ``projector" on physical states $P=\sum_n \bk{n}{n}$ defined by $\bek{\psi_\fin}{P}{\psi_\ini}=\bk{W}{\bar\psi_\fin \otimes \psi_\ini}$ projects on a single state, say $\ket0$, which can be identified as the Hartle-Hawking ``wave function of the universe" defined by the so called ``no boundary proposal"  \cite{Hawking:1980gf}.   We do not expect this factorization to survive higher orders, where the projector can regain its general form. 

From the point of view of cosmology, the system we have described opens in principle the way to the description of inhomogeneous degrees of freedom at the bounce, circumventing the difficulties of the model given in \cite{Rovelli:2008dx}. In particular, the covariant dynamics used here can readily be extended to larger graphs.  Coherent states have been largely used in loop quantum cosmology (see for instance \cite{Ashtekar:2006rx,Ashtekar:2006uz,Ashtekar:2006wn,Ashtekar:2006es}) in particular in relation to the problem of finding effective equations or in numerical simulations \cite{Singh:2005xg,Bojowald:2009jk,Bojowald:2009zz}.  Here, however, homogeneous and isotropic states appear naturally as states peaked on homogeneous and isotropic \emph{mean values} of the quantum states, in the context of a formalism which --we stress-- is not a reduction of the dynamics to homogeneous and isotropic degrees of freedom.  In physical terms, these states represent a universe where inhomogeneous and anisotropic degrees of freedom are taken into account but fluctuate around zero. This provides also an elegant solution of the problem of having to choose between coordinate or momenta in imposing a symmetry reduction in cosmology \cite{Engle:2005ca,Engle:2007zz,Engle:2007qh}.  Ideally, this formalism could describe inhomogeneous and anisotropic quantum fluctuations of the geometry at the bounce. 


\subparagraph{Acknolegements}
A warm thank to Antonino Marcian\`o, Elena Magliaro and Claudio Perini, for a continuous collaboration that has been essential for the developments of the ideas in this work.  We also thank the partecipants to the workshop ``Open problems in Loop Quantum Gravity'' in Zakopane, Poland, for many interesting comments. The work of E.B. is supported by a Marie Curie Intra-European Fellowship within the 7th European Community Framework Programme.

\bibliographystyle{hunsrt}
\bibliography{quantum_cosmology,states,all,biblio}
\end{document}